\documentclass[prl,amsmath,amssymb,groupedaddress,superscriptaddress]{revtex4}
\usepackage[dvips]{graphicx}
\usepackage{float,graphicx}
\usepackage{epsfig}

\def\>{\rangle}
\def\<{\langle}

\begin{document}
\title{Emission spectra of atoms with non-Markovian interaction: Fluorescence in a photonic crystal}
\author{In\'es de Vega$^{\ddag}$ and Daniel Alonso$^{\dagger,\gamma}$}
\affiliation{$^{\ddag}$ Max-Planck Institut f\"ur Quantenoptik,Hans-Kopfermann-Str 1, Garching, D-85748, Germany. EU \\
$^{\dagger}$ Departamento de F\'{\i}sica Fundamental
y Experimental, Electr\'onica y Sistemas. Facultad de F\'{\i}sica. Universidad de La Laguna,\\
La Laguna 38203, Tenerife, Spain, EU.\\
$^{\gamma}$ Instituto Universitario de Estudios Avanzados en F\'isica At\'omica, Molecular y Fot\'onica. Facultad de F\'{\i}sica. Universidad de La Laguna,\\
La Laguna 38203, Tenerife, Spain, EU.}
\begin{abstract}
We present a formula to evaluate the spontaneous emission spectra of an atom in contact with a radiation field with non-Markovian effects. This formula is written in terms of a two-time correlation of system observables and the environmental correlation function, and depends on the distance between the emitting atom and the detector. As an example, we use it to analyze the fluorescence spectra of a two level atom placed as an impurity in a photonic crystal. The radiation field within those materials has a gap or discontinuity where electromagnetic modes cannot propagate in the stationary limit. In that situation, the atomic emission occurs in the form of evanescent waves which are detected with less efficiency the farther we place the detector.
The methodology presented in this paper may be useful to study the non-Markovian dynamics of any quantum open system in linear interaction with a harmonic oscillator reservoir and within the weak coupling approximation.

\end{abstract}

\pacs{03.65. Ca, 03.65.Yz, 42.50.Lc}
\maketitle

\section{Introduction}  

Most of the theoretical work developed in the context of quantum open systems relies in the use of the {\it Markov approximation}, which assumes that when a quantum open system (i.e. the atom) interacts with its environment (for instance a radiation field), the later is not affected in a time scale comparable to that of the system. In other words, the radiation field absorbs the emitted photon instantaneously in comparison to the atomic evolution time scale.

However, the Markovian assumption is only valid for radiation fields in which the photonic density of states (DOS) is a continuous and smoothly varying function in the frequency space. When this function varies in a range which is comparable to the atomic spontaneous emission rate $\Gamma$, then the environmental relaxation time (also correlation time) $\tau_c$, can no longer be considered negligible in comparison with the atomic relaxation time. We are then dealing with {\it non-Markovian} interactions, in which the environment does not absorb instantaneously every photon it receives from the atom, but needs a certain memory time $\tau_c$ to do it and return to its equilibrium state.

Non-Markovian interactions may occur when the radiation field is placed in some structured environment, so that its photonic DOS may present sudden jumps or singular behavior in a frequency range comparable to the spontaneous emission rate. For instance, in the so-called {\it photonic crystals (PC)} \cite{Yab87,Joh87}, there is a periodicity in the refraction index which produces the scattering of light outside of the crystal for certain frequencies that are related to the lattice periodicity. Since these modes are absent from the crystal, the photonic DOS is zero in the corresponding frequencies, so that a \textit{gap} is formed. The frequency ranges in which the photonic DOS is non-zero are known as \textit{bands}. Within the bands, the photonic DOS varies in frequency range comparable to the atomic spontaneous emission rate, which gives rise to long memory time for the environment. This kind of non-Markovian interaction may also be found in a quantum cavity, when the range
  of variation of the photonic DOS is comparable to the spontaneous emission rate \cite{QuantumOptics,Pur46,Kle81}.



In non-Markovian interactions the emitted photon may be eventually reabsorbed by the atom, which produces a back-action in the system dynamics, both in the evolution of its expectation values and the multiple-time correlation functions (MTCF) of its observables. An important consequence of this is that some related physical quantities such as the {\it atomic emission spectra} also display non-Markovian features. Hence, it is necessary to obtain a more general formula for the atomic emission spectra without using the Markovian approximation.

In most of the literature (see for instance \cite{QuantumOptics} and references therein), the analytical expression for the spectra is derived by assuming the Markov approximation at some stage. To be more specific, in photodetector experiments the emission spectra depends on a two-time correlation of the positive and negative operators of the radiation field emitted by the atom. In order to evaluate it, this two-time correlation is expressed in terms of a two-time correlation of system observables, which is done by assuming the Markovian approximation. In the end, it is find that within this approximation the spectra is related to a Fourier transform of the system two-time correlation \cite{QuantumOptics,StatisticalMethodsQuantumOptics}. Moreover, once the two-time correlation is computed through the Quantum Regression Theorem, a Lorentzian profile for the spectra is obtained. This theorem states that the evolution of $N$-time correlation functions can be computed with the  master equation by only considering a rotation in its initial condition.

In this paper, we derive a formula to compute the atomic emission spectra for non-Markovian interactions. We show that the two-time correlation of the positive and negative operators of the radiation field is related to a two-time correlation of system observables through a double convolution integral, that also depends on the environmental correlation function. The decaying of this environmental correlation function gives the correlation time or relaxation time of the environment $\tau_c$, which for non-Markovian interactions is non-zero. Moreover, the two-time correlation of the system observables can no longer be described within the so-called Quantum Regression Theorem (QRT), since the derivation of such formula relies in the Markov approximation \cite{Ons31,Lax63,Lax67}. In the non-Markovian case, the set of evolution equations derived in \cite{ADV05,ADV06} should be used to evaluate MTCF. Moreover, the formula we derive for the emission spectra is dependent on the spatial distance between the emitting atom and the detector.

In order to illustrate the theory, we study the fluorescence spectra of a two-level atom in contact with the modified radiation field that exists within a photonic crystal. In the stationary limit, the emission in the frequency region corresponding to the gap is made in the form of evanescent modes, which present an exponentially decaying spatial dependency. It is clear that when evanescent modes are present in the emission, the fluorescence spectra depends dramatically on the distance from the atom to the detector.

The plan of the paper is the following. In Sec. \ref{SecII} we present the general Hamiltonian for the class of systems that we want to describe, as well as the environmental correlation function that characterizes the interaction. In Section \ref{SecIII} we present a formula to obtain a non-Markovian spectra in terms of a two-time correlation of system observables, convoluted with the environmental correlation function. This formula depends explicitly on the distance between the emitting system and the detector, and is one of the main results of this paper. Section \ref{SecIV} is devoted to review the evolution equations of system two-time correlations within the weak coupling limit. In the last part of the paper (Section \ref{SecV}) we apply the theory to compute the fluorescence spectra of a two-level atom placed as an impurity in a photonic crystal, showing that the distance between the atom and the detector is an important parameter to take into account if we want to characterize correctly the spectra in the gap region.

\section{Non-Markovian interactions}

\label{SecII}
Let us consider the model Hamiltonian to study the dynamics of $\cal{S}$ with Hamiltonian $H_S$, in interaction with $H_B$. Assuming a linear coupling, we have
\begin{eqnarray}
H_{tot}&=&H_S+ H_B + H_I \cr
&=& H_S + \sum_\lambda \omega_\lambda a_\lambda^\dagger a_\lambda + \sum_\lambda g_\lambda \left(a_\lambda^\dagger L  + L^\dagger  a_\lambda\right),
\label{chapdos1}
\end{eqnarray}
where the operator $L$ acts on the Hilbert space of the system and $a_\lambda,a_\lambda^\dagger$ are the annihilation and creation operators on the environment Hilbert space. The $g_\lambda's$ are the coupling constants that can be taken as real numbers, and the $\omega_\lambda's$ are the frequencies of the harmonic oscillators that constitute the environment.

Along the paper we will assume the weak coupling approximation, which consists in considering that the system interaction Hamiltonian is $g$ times smaller than the free Hamiltonian, which includes both the Hamiltonian of the system and the environment. We will also  use for simplicity units in which $\hbar= 1$.

The linear model of interaction Hamiltonian is good enough to characterize most of the interactions in quantum optics related to processes in which only one photon is involved, and which can be described within the dipolar approximation.

The dynamics of the quantum mean values and MTCF can be computed in two distinct ways: {\it First}, by evolving vectors on the system Hilbert space (see for instance \cite{ADV05,ADV06}). In this evolution there are two important quantities that describe the action of the environment into the system: the environmental correlation function
\begin{equation}
\alpha(t-\tau)=\sum_\lambda g_\lambda^2 e^{-i \omega_\lambda (t-\tau)},
\label{chapdos15}
\end{equation}
which as noted before decays within a time $\tau_c$ and describes the dissipation of the system, and a certain quantity $z_{t}$, such that $\alpha(t-\tau)={\mathcal M}[z_{t}z^*_{\tau}]$, where ${\mathcal M}[]$ represents a Gaussian average. The function $z_t$ can be interpreted as a Gaussian colored noise, and describes the stochastic effect of the many environmental degrees on the system wave vector \cite{GN99a,Breuer,DS97,devega2005a,DVAG05}. {\it Second}, in the Heisenberg picture, by evolving system operators. In these equations, the action of the environment is described entirely through the correlation function (\ref{chapdos15}).

Note that in the Markov case, $\alpha(t-\tau)\approx \delta(t-\tau)$ (corresponding to $\tau_c \approx 0$), and the quantity $z_t$ may represent a Gaussian white noise.

In this paper, we need to compute some two-time correlations of system observables. We will do it within the Heisenberg picture scheme, by using the system of equations derived in \cite{ADV05,ADV06} in the weak coupling limit.

To compute the environment correlation function, it is necessary to perform the sum appearing in (\ref{chapdos15}), which can in principle be done once the value of $g_\lambda$ is known for the particular interaction. For instance, in the our case of an atom dipolarly coupled with the radiation field, the coupling constants are known, and the quantum number is $\lambda\equiv \{{\bf k},\sigma\}$, where ${\bf k}$ is the quantum number corresponding to the wave vector, and $\sigma$ corresponds to the two polarization modes.

However, there are certain problems where the functional form $g_\lambda$ is not known, for instance atoms in contact with the phonons in a solid. In those cases, it is often useful to express the sum appearing in (\ref{chapdos15}) as
\begin{equation}
\alpha(t-\tau)=\int d\omega J(\omega)e^{-i \omega (t-\tau)},
\label{chapdos16}
\end{equation}
where we have defined the so-called spectral function as $J(\omega)=\pi \sum_\lambda g_\lambda^2 \delta(\omega-\omega_\lambda )$. This spectral function can be approximated with some phenomenological models (see for instance \cite{CL84,QuantumDissipativeSystems}).

The spectral function can be expressed as $J(\omega)=G^2 (\omega)\rho(\omega)$, where $G(\omega)$ is the function $g_\lambda$ in the continuum and $\rho(\omega)$ is the photonic DOS of the environment. Inserting this relation in (\ref{chapdos16}) it is easy to see the important role of the photonic DOS in the correlation function, and particularly in its decaying time $\tau_c$.

Finally, we note that despite the weak coupling approximation here used is only valid for weak and medium couplings, and therefore it might not capture completely the non-Markovian behavior of the system, it does describe some important memory effects in its dynamics. Particularly, the evolution of the system observables depend on their past history up to the time $\tau_c$ that the environment takes to relax back to equilibrium after the interaction.

\section{Atomic emission spectra}
\label{SecIII}
In this section we present the formula necessary to obtain the emission spectra of an atom with non-Markovian interaction. To this end, we follow a well known photo-detection model of experiment, the \textit{gedanken spectrum analyzer} that can be found in \cite{QuantumOptics,AtomPhotonInteractions}, and which provides an operational definition of the spectral profile. In this scheme, we consider that the radiation field emitted by a two-level atom is detected by another two-level atom with frequency $\omega$, which is initially prepared in its ground state $\mid 1\rangle$. The detecting atom has Hamiltonian $H_{\mathcal D} =\omega \sigma_z/2$ , and it is placed inside a shutter, that only opens during a certain observation time $T$, during which it receives the emitted radiation, being eventually excited. The Hamiltonian of the emitting atom (with levels $|1\rangle$ and $|2\rangle$) is given by
\begin{eqnarray}
H_S =-\frac{\omega_{12}}{2} (\sigma_{22}-\sigma_{11})=\frac{\omega_{12}}{2}\sigma_z,
\label{model1a}
\end{eqnarray}
and the total Hamiltonian of emitting atom and radiation field is described by a Hamiltonian $H_R$, given by $H_R=H_S +H_B+\sum_\lambda g_\lambda ( L^{\dagger}a_\lambda +a_\lambda^\dagger L)$.

The Hamiltonian of the total system  (detector atom, emitting atom and radiation field) reads as follows,
\begin{eqnarray}
H=H_{\mathcal D} +H_R +W.
\label{es1}
\end{eqnarray}
Here the coupling between the detecting atom $H_{\mathcal D}$ with the radiation field, with Hamiltonian $H_R$, is dipolar and given by a Hamiltonian $W$, which in interaction image with respect to the detector is given by
\begin{eqnarray}
\tilde{W}(t)=\left[\sigma_{21} {\bf d}^{{\mathcal D}}\cdot {\bf E}^{(+)} ({\bf r},t)e^{i\omega t}+\sigma_{12}{\bf d}^{{\mathcal D}}\cdot{\bf E}^{(-)} ({\bf r},t)e^{-i\omega t}\right],
\label{es2}
\end{eqnarray}
where we have considered  $ d^{{\mathcal D}}_{21}{\bf \hat{d}}^{{\mathcal D}} =d^{{\mathcal D}}_{12}{\bf \hat{d}}^{{\mathcal D}}  =\langle 1 \mid {\bf D} \mid 2\rangle={\bf d}^{{\mathcal D}} $. The super-index ${\mathcal D}$ reminds that these are the components of the detector's dipole, and ${\bf D}$ is the dipolar operator. It is important to note here that the field operators ${\bf E}^{(+)}$ and ${\bf E}^{(-)}$,  correspond to the radiative atoms and the radiation field with $H_R$. The positive part of the field placed in {\bf r} is defined as
\begin{eqnarray}
{\bf E}^{(+)}({\bf r},{\bf r}_0,t)=\sum_\lambda \epsilon_\lambda A_\lambda ({\bf r} ) a_\lambda ({\bf r}_0,t){\bf e}_\lambda
\label{es9chap4}
\end{eqnarray}
and ${\bf E}^{(-)}({\bf r},{\bf r}_0,t)=[{\bf E}^{(+)}({\bf r},{\bf r}_0,t)]^{\dagger}$ \cite{AtomPhotonInteractions}. In the last expression (and from now on) we have added explicitly the dependency on the position ${\bf r}_0$ of the source dipole that originates the field. The quantity $\epsilon_\lambda =\sqrt{\frac{ \omega_\lambda }{2\epsilon_0 \upsilon }}$, with $\upsilon$ the quantization volume. In terms of the coupling strengths we find that $g_\lambda \equiv g_\lambda ({\bf r})=\epsilon_\lambda A_\lambda ({\bf r} ){\bf d}\cdot {{\bf e}}_\lambda $.

As a condition, the time of observation $T$ is much bigger that the inverse of the natural width $\Gamma$ of the detecting atom excited level. In addition, the Rabi frequency of the emitting atom has to be bigger than the inverse of $T$. With this set up, the idea is to calculate the spectral distribution of the fluorescence light, $P(\omega,T)$. This is defined as the probability of excitation of the detecting atom at the time of observation $T$, i.e.
\begin{eqnarray}
P(\omega,T)=Tr_{R,{\cal D}}\left(\mid 2\rangle\langle 2\mid \rho(T)\right),
\label{es3}
\end{eqnarray}
where $\rho (T)$ is the density matrix of the total system at time $T$. In the interaction representation, this density matrix is,
\begin{eqnarray}
\rho(T)=\rho(0)-i\int^{T}_0 \left[\tilde{W}(t),\rho(0)\right]-\int^{T}_0 dt\int^{t}_0 dt' \left[\tilde{W}(t),\left[\tilde{W}(t'),\rho(0)\right]\right],
\label{es4}
\end{eqnarray}
for an expansion in which $\rho(T)\approx \rho(0)$.

Replacing (\ref{es4}) in (\ref{es3}), we get the following expression for $P(\omega,T)$,
\begin{eqnarray}
&&P(\omega,T)= Tr_{R} \left(\int^{T}_0 dt \int^{t}_0 dt' e^{i\omega (t-t')}{\bf d}^{{\mathcal D}}\cdot {\bf E}^{(+)} ({\bf r},{\bf r}_0,t)\rho_R {\bf d}^{{\mathcal D}}\cdot {\bf E}^{(-)} ({\bf r},{\bf r}_0,t')\right.\nonumber\\
&+& \left.\int^{T}_0 dt \int^{t}_0 dt' e^{i\omega (t'-t)}{\bf d}^{{\mathcal D}}\cdot {\bf E}^{(+)} ({\bf r},{\bf r}_0,t')\rho_R {\bf d}^{{\mathcal D}}\cdot {\bf E}^{(-)} ({\bf r},{\bf r}_0,t)\right).
\label{es5}
\end{eqnarray}
If we now change $t$ into $t'$, and $t'$ into $t$ in the second integral,
\begin{eqnarray}
&&P(\omega,T)= Tr_{R} \left(\int^{T}_0 dt \int^{t}_0 dt' e^{i\omega (t-t')}{\bf d}^{{\mathcal D}}\cdot {\bf E}^{(+)} ({\bf r},{\bf r}_0,t)\rho_R {\bf d}^{{\mathcal D}}\cdot {\bf E}^{(-)} ({\bf r},{\bf r}_0,t')\right.\nonumber\\
&+& \left.\int^{T}_0 dt' \int^{t'}_0 dt e^{i\omega (t-t')}{\bf d}^{{\mathcal D}}\cdot {\bf E}^{(+)} ({\bf r},{\bf r}_0,t)\rho_R {\bf d}^{{\mathcal D}}\cdot {\bf E}^{(-)} ({\bf r},{\bf r}_0,t')\right),
\label{es5b}
\end{eqnarray}
both integrand becomes the same. On the other hand, the regions of integration are the complementary parts of a square of side $T$, so that the last expression can be written as
\begin{eqnarray}
P(\omega,T)= \int^{T}_0 dt \int^{T}_0 dt' e^{i\omega (t-t')}g^{(1)}({\bf r},{\bf r}_0;t,t'),
\label{es6}
\end{eqnarray}
where the average $\langle \cdots\rangle=Tr_R \left(\rho_R \cdots\right)$, and we have defined
\begin{eqnarray}
g^{(1)}({\bf r},{\bf r}_0;t,t')=\langle {\bf d}^{{\mathcal D}}\cdot {\bf E}^{(-)} ({\bf r},{\bf r}_0,t){\bf d}^{{\mathcal D}}\cdot {\bf E}^{(+)} ({\bf r},{\bf r}_0,t')\rangle,
\label{firstorderchap4}
\end{eqnarray}
as the first order correlation of the projection of the emitted field in the direction of the dipole.




In order to express the last equation in terms of observables of the emitting atoms, it is necessary to express the field operator, which is defined in (\ref{es9chap4}) in terms of the atomic dipolar momentum. We insert in (\ref{es9chap4}) the solution of the Heisenberg equation for $ a_\lambda ({\bf r}_0,t)$ with the total Hamiltonian (\ref{es1}) , i.e. $d a_\lambda ({\bf r}_0,t)/dt=-i[H , a_\lambda ({\bf r}_0,t)]$,
\begin{eqnarray}
a_{n}({\bf r}_0,t)=a_\lambda ({\bf r}_0,0)e^{-i\omega_\lambda \tau}-i  \int^{t}_0 d\tau g_\lambda  L(t')e^{-i\omega_\lambda (t-\tau)}
\label{es10}
\end{eqnarray}
Here, the coupling constant between emitting atoms and surrounding field is $g_\lambda \equiv g_\lambda ({\bf r}_0)=\epsilon_\lambda A_\lambda ({\bf r}_0){\bf d}\cdot {\bf e}_\lambda $.  After this replacement, the component of ${\bf E}^{(+)}$ in the direction and position of the detector is
\begin{eqnarray}
{\bf d}^{{\mathcal D}}\cdot {\bf E}^{(+)}({\bf r},{\bf r}_0,t)=-i\sum_\lambda g_\lambda  g^{{\mathcal D}}_\lambda  \int^{t}_0 d\tau L(\tau)e^{-i\omega_\lambda (t-\tau )},
\label{es11}
\end{eqnarray}
where we have defined $g^{{\mathcal D}}_\lambda \equiv g^{{\mathcal D}}_\lambda ({\bf r})=\epsilon_\lambda A_\lambda ({\bf r}){\bf d}^{{\mathcal D}}\cdot {\bf e}_\lambda $ as the coupling constant of the detector with the emitted field. In expression (\ref{es11}) we only keep the part of the field corresponding to the radiated field from the dipole (also ``source field''). The other part, corresponding to the quantum vacuum field
\begin{eqnarray}
{\bf E}_0^{(+)}({\bf r},{\bf r}_0,t)=\sum_\lambda \epsilon_\lambda A_\lambda ({\bf r}) a_\lambda ({\bf r}_0 ,0)e^{-i\omega_\lambda t} {\bf e}_\lambda,
\end{eqnarray}
 does not contribute to photo-detection signals, since provided that the field is in the vacuum state $|0\rangle$,
\begin{eqnarray}\langle 0|\left({\bf d}^{{\mathcal D}}\cdot {\bf E}_0^{(-)}({\bf r},{\bf r}_0,\tau)\right)&=&0 \nonumber\\
\left({\bf d}^{{\mathcal D}}\cdot {\bf E}_0^{(+)} ({\bf r},,{\bf r}_0,0)\right)|0\rangle&=&0
\end{eqnarray}
in (\ref{firstorderchap4}).
We now introduce a distribution function $\rho({\bf r}_0)$, which describes the density of radiating atoms at a given positions of the crystal. Assuming that the atomic distribution is the same for each Wigner Seitz cell (WSC) of the crystal \cite{AM}, we perform an average over
the atomic distribution within the crystal, obtaining the following expression for the positive part of the radiation field,
\begin{eqnarray}
\langle {\bf d}^{{\mathcal D}}\cdot{\bf E}^{(+)}({\bf r},{\bf r}_0,t)\rangle_{{\bf r}_0 } =\frac{1}{N_e }\int_{WSC} d{\bf r}_0 \rho({\bf r}_0) {\bf d}^{{\mathcal D}}\cdot{\bf E}^{(+)}({\bf r},{\bf r}_0,t)
\end{eqnarray}
where $N_e =\int_{WSC} d{\bf r}_0 \rho({\bf r}_0)$ is the number of active atoms in a cell. Considering one single atom at position ${\bf r}_a$ in each cell, such that $\rho({\bf r}_0)=\delta({\bf r}_0-{\bf r}_a)$ and $N_e =1$. Then we have

\begin{eqnarray}
\langle{\bf \hat{d}}_D \cdot {\bf E}^{(+)}({\bf r},{\bf r}_0,t)\rangle_{{\bf r}_0 }=-i \int^{t}_0 d\tau {\cal S}({\bf r},{\bf r}_a,t,\tau) L(\tau)
\label{es112}
\end{eqnarray}
with the
\begin{eqnarray}
{\cal S}({\bf r},{\bf r}_a,t,\tau)=\sum_\lambda  g^{{\mathcal D}}_\lambda g_\lambda e^{-i\omega_\lambda (t-\tau)}e^{i({\bf r}-{\bf r}_a ){\bf k}},
\label{chapdosd}
\end{eqnarray}
where for simplicity in the notation we set $g_\lambda \equiv g_\lambda ({\bf r}_a)$.
 Replacing (\ref{es112}) in the integrand of (\ref{es6}), we get
\begin{eqnarray}
g^{(1)}({\bf r},{\bf r}_a,t,t')&=&\langle g^{(1)}({\bf r},{\bf r}_0,t,t')\rangle_{{\bf r}_0}\nonumber\\
&=&\bigg\langle \bigg(\langle {\bf d}^{{\mathcal D}}\cdot {\bf E}^{(-)} ({\bf r},{\bf r}_a,t)\rangle_{{\bf r}_0}\bigg)\bigg(\langle {\bf d}^{{\mathcal D}}\cdot {\bf E}^{(+)} ({\bf r},{\bf r}_a,t')\rangle_{{\bf r}_0}\bigg)\bigg\rangle \nonumber\\
&=&\int^t_0 d\tau \int^{t'}_0 d\tau' {\cal S}^* ({\bf r},{\bf r}_a,t,\tau) {\cal S}({\bf r},{\bf r}_a,t',\tau')\langle L^{\dagger}(\tau)L(\tau')\rangle
\end{eqnarray}
Following (\ref{es6}) the spectra has the following form,
\begin{eqnarray}
P(\omega,T)&=&\int^{T}_0 dt \int^{T}_0 dt' e^{i\omega (t-t')}g^{(1)}({\bf r},{\bf r}_a,t,t')\nonumber\\
&=&\int^{T}_0 dt \int^{T}_0 dt' e^{i\omega (t-t')}\bigg\{\int^t_0 d\tau \int^{t'}_0 d\tau' {\cal S}^* ({\bf r},{\bf r}_a,t,\tau) {\cal S}({\bf r},{\bf r}_a,t',\tau') \nonumber\\
&&\times\langle L^{\dagger}(\tau)L(\tau')\rangle \bigg\}.
\label{es}
\end{eqnarray}
In order to compute the fluorescence spectra, the limit of $T\rightarrow \infty$ has to be taken, so that the signal is observed in the stationary limit. In that case, the above formula corresponds to two Laplace transform of a convolution,
\begin{eqnarray}
P(\omega)= {\cal S}^* ({\bf r},{\bf r}_a,-\omega) {\cal S}({\bf r},{\bf r}_a,\omega)\langle L^{\dagger}(-\omega)L(\omega)\rangle .
\label{pesp}
\end{eqnarray}
Since
\begin{eqnarray}
P(\omega,\infty)=P(\omega)=Tr_{R,{\cal D}}\left(\mid 2\rangle\langle 2\mid \rho(\infty)\right)
\end{eqnarray}
we are dealing with the detector reduced density matrix in the stationary limit.

\subsection{Special case: No spatial dependency}
When there is no spatial dependency of the functions ${\cal S}^* ({\bf r},{\bf r}_a,t,\tau)$, i.e.
\begin{equation}
e^{i{\bf k}\cdot ({\bf r}_0-{\bf r}_a )}\approx 1,
\end{equation}
then
\begin{eqnarray}
{\cal S}^* ({\bf r},{\bf r}_a,t,\tau)=\alpha(t-\tau)
\end{eqnarray}
they are the usual correlation functions $\alpha(t-\tau)$. In that case, we can write (\ref{es}) as
\begin{eqnarray}
P(\omega,T)&=&\int^{T}_0 dt \int^{T}_0 dt' e^{i\omega (t-t')}\bigg\{\int^t_0 d\tau \int^{t'}_0 d\tau' \alpha^* (t-\tau) \alpha(t'-\tau') \langle L^{\dagger}(\tau)L(\tau')\rangle \bigg\}.
\label{ess}
\end{eqnarray}
Hence, the fluorescence spectra can be written as
\begin{eqnarray}
P(\omega)=\alpha^* (-\omega) \alpha(\omega)\langle L^{\dagger}(-\omega)L(\omega)\rangle .
\label{pnoesp}
\end{eqnarray}

In addition, in the Markov case, such correlation is a delta function, $\alpha(t-\tau)=\Gamma \delta(t-\tau)$, and the last formula is just
\begin{eqnarray}
P(\omega,T)=\Gamma^2 \int^{T}_0 dt \int^{T}_0 dt' e^{i\omega (t-t')}  \langle L^{\dagger}(t)L(t')\rangle,
\label{esM1}
\end{eqnarray}
which in the equilibrium, and with the observation time $T\rightarrow\infty$, gives rise to the usual expression for the power spectra,
\begin{eqnarray}
P(\omega)=\Gamma^2 \int^{\infty}_0 d\tau e^{i\omega \tau} \langle L^{\dagger}(0)L(\tau)\rangle.
\label{esM2}
\end{eqnarray}
The system correlations $\langle L^{\dagger}(0)L(\tau)\rangle$ can be computed with the Quantum Regression Theorem.

\section{Two-time correlation function of system observables in the weak coupling limit}
\label{SecIV}

In order to evaluate the spectra from eq.(\ref{pesp}), it is necessary to obtain the evolution of two-time correlations of the system observables. A theory to compute multiple time correlation functions in the non-Markovian regime has been developed in \cite{ADV05,ADV06}. In this section, we briefly present some of the main results, and the formula to compute two-time correlations in the weak coupling limit needed in eq.(\ref{pesp}).

The evolution equation of the quantum mean value of $A$, provided that the total initial state is $\mid\Psi_0 \rangle=\mid\psi_0\rangle\mid 0\rangle$, is equal to
\begin{eqnarray}
&&\frac{d}{dt_1}\langle\Psi_0 \mid A(t_1)\mid\Psi_0 \rangle=i\langle\Psi_0 \mid  [H_S (t_1) ,A(t_1)]\mid\Psi_0 \rangle \nonumber\\
&+& \int_0^{t_1} d\tau \alpha(t_1-\tau)\langle\Psi_0 \mid [L^{\dagger}(t_1),A(t_1)] L(\tau)\mid\Psi_0\rangle\nonumber\\
&+& \int_0^{t_1} d\tau \alpha^*(t_1-\tau)\langle\Psi_0 \mid L^{\dagger}(\tau)[A(t_1),L(t_1)]\mid\Psi_0\rangle.
\label{eq29}
\end{eqnarray}

In the same manner the evolution of the quantum mean value $\langle A(t_1 )B(t_2 )\rangle$ is given by
\begin{eqnarray}
&&\frac{d\langle \Psi_0 \mid A(t_1 )B(t_2 )\mid\Psi_0\rangle}{dt_1 }=i\langle \Psi_0 \mid [H_S (t_1 ),A(t_1 )]B(t_2 )\mid\Psi_0\rangle \nonumber \\
&+&\int_0^{t_1 }d\tau \alpha^*(t_1 -\tau) \langle \Psi_0 \mid L^{\dagger}(\tau)[A(t_1 ),L(t_1 )]B(t_2 )\mid\Psi_0\rangle\nonumber\\
&+&\int_{t_2 }^{t_1 }d\tau \alpha(t_1 -\tau)\langle\Psi_0 \mid [L^{\dagger}(t_1 ),A(t_1 )]L(\tau)B(t_2 )\mid\Psi_0\rangle\nonumber\\
&+&\int_0^{t_2 } d\tau  \alpha(t_1 -\tau)\langle\Psi_0 \mid [L^{\dagger}(t_1 ),A(t_1 )]B(t_2 )L(\tau)\mid\Psi_0\rangle .
\label{eq45}
\end{eqnarray}

Equations (\ref{eq29}) and (\ref{eq45}) represent the evolution of quantum mean values and two-time correlations respectively, obtained \textit{without the use of any approximation}. However, it is clear that these equations are open, in the sense that quantum mean values depend on two-time correlations, while two-time correlations depend on three time correlations. In general when no approximations are made, $N$-time correlation depend on $N+1$-time correlations, what gives rise to a hierarchy structure of MTCF as described in \cite{DVA06b}.

For practical computations, it is necessary to make some assumptions. In our case we perform a perturbative expansion on the parameter $g$, so that the former equations become closed. Particularly, the perturbative hypothesis is applied to the operators $L(\tau)$ and $L^{\dagger}(\tau)$ that appears in equations (\ref{eq29}) and (\ref{eq45}). The idea is to transform groups of operators with a two-time dependency on groups with a one time-dependency \footnote{We stress that by time dependency we mean that related to the evolution operator ${\cal U}_I (t)$ that represents the evolution with the total Hamiltonian, in the interaction image.}. For instance, the first dissipative term in (\ref{eq29}) and (\ref{eq45}) is of type
\begin{eqnarray}
&&L^{\dagger}(\tau)\left\{[L,A]\right\}(t_i)={\mathcal U}^{-1}_I (t_i 0)L(\tau,t_i)[L,A]{\mathcal U}_I (t_i 0 )\nonumber\\
&=&\left\{V_{\tau-t_i}L^{\dagger}[L,A]\right\}(t_i)+{\mathcal O}(g).
\label{eq61}
\end{eqnarray}
Here, we have used a Liouville super-operator $V_t$ which acts only on the next system operator as $V_t L =e^{iH_S t}L e^{-i H_S t}$.
In the same way, the second dissipative term in (\ref{eq45}) can be expanded as
\begin{eqnarray}
&&L(\tau)B(t_{i+1})={\mathcal U}^{-1}_I (t_{i+1} 0)L(\tau,t_{i+1})B{\mathcal U}_I (t_{i+1} 0 )\nonumber\\
&=&\left\{V_{\tau-t_{i+1}} LB\right\}(t_{i+1})+{\mathcal O}(g),
\label{eq62}
\end{eqnarray}
up to zero order. Note that just as before, the dissipative terms are already of second order due to the appearance of the second order function $\alpha(t)$, so that only zero order expansions of the rest of the quantities are needed. Then, up to second order in $g$, we obtain the following equation for quantum mean values,
\begin{eqnarray}
&&\frac{d}{dt_1}\langle\Psi_{0}\mid A(t_1)\mid\Psi_{0}\rangle=
i \langle\Psi_{0}\mid \left\{[H_S ,A]\right\}(t_1)\mid\Psi_{0}\rangle\nonumber\\
&+&\int^{t_1}_0 d\tau \alpha^*(t_1-\tau)\langle\Psi_{0}\mid \left\{V_{\tau-t_1} L^{\dagger}[A,L]\right\}(t_1)\mid\Psi_{0}\rangle\nonumber\\
&+&\int^{t_1}_0 d\tau \alpha(t_1-\tau)\langle\Psi_{0}\mid \left\{[L^{\dagger},A]V_{\tau-t_1} L \right\}(t_1)\mid\psi_{0}\rangle\nonumber\\
+{\mathcal{O}}(g^3),
\label{qrt4}
\end{eqnarray}
and for two-time correlations
\begin{eqnarray}
&&\frac{d}{dt_1}\langle\Psi_{0}\mid A(t_1)B(t_2)\mid\Psi_{0}\rangle=
i \langle\Psi_{0}\mid \left\{[H_S ,A]\right\}(t_1)B(t_2)\mid\Psi_{0}\rangle\nonumber\\
&+&\int^{t_1}_0 d\tau \alpha^*(t_1-\tau)\langle\Psi_{0}\mid \left\{V_{\tau-t_1} L^{\dagger}[A,L]\right\}(t_1)B(t_2)\mid\Psi_{0}\rangle\nonumber\\
&+&\int^{t_1}_0 d\tau \alpha(t_1-\tau)\langle\Psi_{0}\mid \left\{[L^{\dagger},A]V_{\tau-t_1} L \right\}(t_1)B(t_2)\mid\psi_{0}\rangle\nonumber\\
&+&\int_0^{t_2} d\tau\alpha(t_1-\tau) \langle\Psi_{0}\mid \left\{[L^{\dagger}, A]\right\}(t_1)\left\{[B,V_{\tau-t_2 } L]\right\}(t_2)\mid\Psi_{0}\rangle+{\mathcal{O}}(g^3).
\label{qrt5}
\end{eqnarray}
While the first two terms of (\ref{qrt5}) are analogous to those of (\ref{qrt4}), the equation for two-time correlations contains an additional term that does not appear in the evolution of quantum mean values. Since the Quantum Regression Theorem (QRT) states that the evolution of two-time correlations has the same coefficients as the evolution of one-time correlations, these equations show that the theorem is not satisfied for non-Markovian interactions. The validity of the QRT have been previously discussed by some authors \cite{Ford1996,Lax2000,Ford2000a,Ford2000b}. Notice that for Markovian interactions the last term of (\ref{qrt5}) vanishes, since the corresponding correlation function $\alpha(t_1-\tau)=\Gamma \delta(t_1-\tau)$ is zero in the domain of integration from $0$ to $t_2$, so that the QRT is valid in this case. Note that an equation similar to (\ref{qrt5}) can be derived for initial condition $\langle A(t_1)B(t_1)\rangle$, and for $t_2>t_1$. For the Markov case, it can also be shown that this equation is equivalent to the one given by the QRT for such an initial condition. For our purposes, we shall use equation (\ref{qrt5}) to evaluate the evolution of non-Markovian two-time correlations.

\section{Fluorescence spectra of a two level atom}
\label{SecV}

The formulas derived up to now are valid for the emission spectra of an arbitrary atom, provided that the interaction Hamiltonian has the linear form (\ref{chapdos1}) and that the weak coupling approximation is valid.
In this section, we show how to express the Hamiltonian for a driven two-level  atom, so that a particular form for the coupling operator $L$ is given.

 As in \cite{FS64}, we consider the interaction Hamiltonian $H_{SL}$ of the atom with the classical coherent monochromatic laser field in the usual rotating wave approximation form \cite{QuantumOptics}:
\begin{equation}
H_{SL}=\epsilon(\sigma_{21}e^{-i(\omega_{L}t+\phi_{T})}+\sigma_{12}e^{i(\omega_{L}t+\phi_{T})}),
\end{equation}
where $\omega_{L}$ is the frequency of the laser in the coherent state $\alpha e^{-i\omega_{L}t}$, with $\alpha=|\alpha| e^{-i\phi_{L}}$. The quantity $\epsilon={\bf d_{21}}\varepsilon$ is the Rabi frequency, and $|\varepsilon|=\sqrt{\omega_{L} /2\epsilon_{0}\upsilon}\sqrt{\bar{N}_{L}}e_{L}$ the laser field magnitude, where $\upsilon$ is the volume of the cavity, and $\bar{N}_{L}=\mid \alpha^2 \mid$ and $e_{L}$, are respectively the mean number of photons and the polarization of the laser mode. The phase $\phi_{T}$ is a global phase defined as $\phi_{T}=\phi_{L}-\pi/2$. Because of the magnitude of the laser field, the Hamiltonian $H_{SL}$ should be considered as part of the non-interacting Hamiltonian $H_{0}$, where in this case $H_{0}=H_S+H_{B}+H_{SL}$.
We can eliminate the explicit dependence on the laser frequency by changing to a rotating frame with a frequency $\omega_{L}$ by using the unitary operator
\begin{equation}
U_t=e^{\left[i{\omega_L}t+i\phi_{T}\right]\left[\sum_{\lambda} a^{\dagger}_{\lambda}a_{\lambda}+(\sigma_{22}-\sigma_{11})\right]}.
\end{equation}

If $\Delta_\lambda=\omega_\lambda -\omega_L$ and $\Delta_{SL}=\omega_S -\omega_L$ the rotated Hamiltonian $H'=H_{0}'+H_I'$, where
\begin{equation}
H_{0}'=\sum_{\lambda}\Delta_\lambda a_{\lambda}^{\dagger}a_{\lambda}+\frac{1}{2}\omega_{S}\sigma_{3}+\epsilon[\sigma_{21}+\sigma_{12}]
\end{equation}
and
\begin{equation}
H_I'=i\sum_{\lambda}g_{\lambda}(\sigma_{12} a_{\lambda}^{\dagger}-a_{\lambda}\sigma_{21}),
\end{equation}
can still be expressed in a simpler way by projecting it into the dressed atomic basis. The new Hamiltonian $\tilde{H}=V^{-1}H'V$, where
\begin{eqnarray}
V=\left(
\begin{array}{cc}
{\bf c}&-{\bf s} \\
{\bf s}&{\bf c}
\end{array}
\right),
\label{transfo}
\end{eqnarray}
will be of the form (\ref{chapdos1}). The constants appearing in the unitary transformation matrix $V$ are ${\bf c}=\cos\phi$ and ${\bf s}=\sin\phi$, where the angle, $\phi$, is given by $\sin^{2} \phi=\frac{1}{2} [1-sgn(\Delta_{SL})/\sqrt{\epsilon^{2}/\Delta^{2}_{SL})+1}]$. The non interacting dressed state Hamiltonian $\tilde{H}_{0}=\tilde{H}_{S}+\tilde{H}_{B}+\tilde{H}_{SL}$ is equal to,
\begin{eqnarray}
\tilde{H}_{0}=\Omega R_{3}+ \sum_{\lambda}\Delta_{\lambda}a^{\dagger}_\lambda a_{\lambda},
\end{eqnarray}
and the interaction Hamiltonian $\tilde{H}_{I}$ has the same form as in (\ref{chapdos1}) once the interaction operator $L$ is defined as:
\begin{equation}
L={\bf cs}R_{3}+{\bf c}^{2}R_{12}-{\bf s}^{2}R_{21}.
\label{Llaser}
\end{equation}
Here, $R_{ij}=|\tilde{i}\rangle\langle \tilde{j}|$ are the atomic operators defined in the dressed state basis $\{|\tilde{1}\rangle,|\tilde{2}\rangle\}$, and $R_{3}=R_{22}-R_{11}$. The quantity $\Omega=[\epsilon^2+\Delta_{SL}^2/4]^{1/2}$ is called the generalized Rabi frequency.

According to (\ref{transfo}), the relation between the dressed atomic operators and the bare atomic operators is the following,
\begin{eqnarray}
\sigma_{12}={\bf cs}R_{3}+{\bf c}^{2}R_{12}-{\bf s}^2 R_{21}\nonumber\\
\sigma_{21}={\bf cs}R_{3}-{\bf s}^2 R_{12}+{\bf c}^2 R_{21}\nonumber\\
\sigma_{3}=({\bf c}^2 -{\bf s}^2)R_{3}-2{\bf cs}(R_{12}+R_{21}).
\label{transfo2}
\end{eqnarray}
It can be easily verified that the case of spontaneous emission corresponds to switching off the laser, and therefore $L=\sigma_{12}$.

Once the coupling operator $L$ is known, the two time correlation $\langle L^\dagger(\tau)L (\tau')\rangle$ can be easily computed with equation (\ref{qrt5}). Now, in order to obtain the emission spectra with (\ref{pesp}), it is necessary to derive the Laplace transform of the spatial dependent correlation function,  ${\cal S}^* ({\bf r},{\bf r}_a,\omega)$. Notice that the interaction with the field enters into the formula (\ref{pesp}) in two ways:
\begin{itemize}
\item Indirectly, through the correlation function $\alpha(t_1-\tau)$ entering in equation (\ref{qrt5}).
\item Directly, through the function ${\cal S}^* ({\bf r},{\bf r}_a,\omega)$.
\end{itemize}
Let us make here a remark. In order to compute the atomic dynamical equations, we assume that the field is composed of extended modes, in the sense that they correspond to a real wave vector. Hence, the spatial dependency is assumed to be $\exp{(i {\bf k}\cdot{\bf r})}\approx 1$ within the optical region.
This assumption that the field is composed only of extended modes corresponds to consider the photonic crystal as an infinite and perfect structure, so that evanescent modes are in principle not present. In other words, even if evanescent modes are solutions of the eigenvalue problem, normally they cannot be excited inside a perfect (infinite and without defects) photonic crystal. This is because the do not satisfy the translational simmetry boundary condition of the crystal.

Nevertheless, a defect placed in the crystal (in the our case an emitting atom) may substain such modes.
To be more specific, once an impurity atom with resonant frequency within the gap region is placed within the crystal, its emission will be in the form of evanescent modes (i.e. with imaginary wave vector). Then, the approximation $\exp{(i {\bf k}\cdot{\bf r})}\approx 1$ is no longer valid for such field. As will be explained in detail later, this fact is very relevant to compute the emission spectra, which appears to be strongly dependent on the atom-detector distance through the spatially dependent function ${\cal S}^* ({\bf r},{\bf r}_a,\omega)$.

In the next section, we present the calculus of $\alpha(t-\tau)$ and  ${\cal S}^* ({\bf r},{\bf r}_a,\omega)$ necessary to compute the system fluctuations and the fluorescence spectra respectively. Both quantities depend on the particular interaction considered. We study them for the case of an atom interacting with the radiation field within a photonic crystal.

\section{Environment correlation functions for an atom in a  PBG}
\label{thecorrelation}

In this section, we calculate the correlation function characteristic of the dipolar coupling of a two level atom to the modified radiation field within a PC. In that case, the coupling constants entering in (\ref{chapdos15}) are
\begin{eqnarray}
g_{\lambda}=-i \sqrt{\frac{1}{2\hbar\epsilon_0 \omega_\lambda }}\omega_{12}\hat{\bf e}_{{\bf k}\sigma} \cdot {\bf d}_{12} e^{i{\bf k}\cdot {\bf r}},
\label{model3}
\end{eqnarray}
 where ${\bf d}_{12}$ is the atom dipolar moment, $\hat{\bf e}_{{\bf k}\sigma}$ is the unit vector in the direction of the polarization $\sigma$ for a given wave vector ${\bf k}$, and $\epsilon_0$ is the electric vacuum permittivity. In the last expression we have already expanded the potential vector in terms of plane waves. This is also valid for radiation fields in photonic crystals, but with the additional requirement that these plane waves are periodic, i.e. satisfy the Bloch theorem.  As we have already noted in the introduction, the presence of a periodic dielectric structure produces the appearance of optical bands, which are themselves periodic structures in the reciprocal lattice. Periodic eigenvalues (the frequencies), correspond to periodic eigenvectors (the field). However, although the quantity $\lambda \equiv\{{\bf k},\sigma,n\}$, where $n$ denotes each band, this index will be here dropped, since we consider the dynamics of the two level system with rotating frequency nearby (or inside) a single band.

In order to calculate $\alpha(t)$, and ${\mathcal S}({\bf r},{\bf r}_a,t)$, the sums appearing in equations (\ref{chapdos15}) and (\ref{chapdosd}) respectively, have to be performed, by using the functions $g_{\lambda}$ as defined in (\ref{model3}).
In that way we have
\begin{eqnarray}
\alpha(\tau)=\gamma(\frac{a}{2 \pi})^3 \sum_{\sigma}\int_{1BZ}d{\bf k}\frac{ |\hat{e}_{{\bf k},\sigma}\cdot \hat{u}_d|
^{2}}{\omega({\bf k})}e^{-i\omega({\bf k})\tau},
\label{generalcorr5}
\end{eqnarray}
and
\begin{eqnarray}
{\cal S}({\bf r},{\bf r}_a,\tau)=\gamma(\frac{a}{2 \pi})^3 \sum_{\sigma}\int_{1BZ}d{\bf k} \frac{ |\hat{e}_{{\bf k},\sigma}\cdot \hat{u}_d||\hat{e}_{{\bf k},\sigma}\cdot \hat{u}^{\cal D}_d|}{\omega({\bf k})} e^{-i\omega({\bf k})\tau}e^{i({\bf r}-{\bf r}_a ){\bf k}},
\label{chapdosd2}
\end{eqnarray}
where $\hat{u}^{\cal D}_d$ is the unitary vector corresponding to the dipolar moment of the detector, and the integrals are performed over the first Brillouin zones of the crystal. Also, it is clear from (\ref{model3}) that the constant $\gamma=\omega_{12}^2 d^2_{12} /(2\epsilon_0 \hbar)$. In order to calculate the integrals appearing in (\ref{generalcorr5}) and (\ref{chapdosd2}), it is clear that the dispersion relation $\omega({\bf k})$ is needed.

The dynamics near the edge of the band are often described through an effective mass approximation to the full dispersion relation, based on its expansion in the vicinity of the band edge (see for instance \cite{FS64,SQ95,SW90,SW91}). Within this approximation, and considering the simplest case of a cubic lattice, the dispersion relation has the parabolic form $\omega({\bf k})=\omega_{c}+{\mathcal A}({\bf k}-{\bf k_0})^{2}$, where ${\bf k_0}$ is the origin of the first Brillouin zone of the crystal (which is the unitary cell in ${\bf k}$ space) about which we perform the expansion in each direction, $\omega_c$ is the frequency of the band edge, and ${\mathcal A}$ is a constant that depends on the specific photonic crystal considered. Although it is valid for the description of several important physical phenomena occurring in PBG materials \cite{Joh87,FS64,PRA64_4,PRA64_6,PRA64_7}, this model of dispersion relation becomes inaccurate for a good description of the short time dynamics. The dispersion relation proposed in \cite{DVAG05} overcomes this problem by reproducing the periodic structure of the bands,
\begin{eqnarray}
\omega({\bf k})=A+\frac{B}{3}\big(\cos(k_{x}a)+\cos(k_{y}a)+\cos(k_{z}a)\big).
\label{waniso}
\end{eqnarray}
Notice that this function reduces to the parabolic model for ${\bf k}\approx {\bf k_0}$.
For a three dimensional band structure with the periodic dispersion relation (\ref{waniso}), the correlation function can be obtained from the equation (\ref{generalcorr5}) once it has been assumed that the function $|\hat{e}_{k,\sigma}\cdot \hat{u}_d|^{2}/\omega({\bf k})$ is a slowly varying function in the $1BZ$. The resulting integral is analytical, and gives rise to the following result:
\begin{eqnarray}
\alpha_{3D}(\tau)&\cong&\gamma\sum_{\sigma}\frac{ |\hat{e}_{{\bf k}_0 ,\sigma}\cdot \hat{u}_d|^{2}}{\omega({\bf k_{0}})}e^{-iA\tau}J_{0}^{3}(\frac{B\tau}{3})\nonumber\\
&\cong&g_{3D}^{2}e^{-iA\tau}J_{0}^{3}(\frac{B\tau}{3}),
\label{caniso1}
\end{eqnarray}
where $J_0$ is the zero order Bessel function.

Let us now consider the spatial dependent correlation function, ${\cal S}({\bf r},{\bf r}_a,t,\tau)$. As before, we assume that the quantity $\frac{ |\hat{e}_{{\bf k},\sigma}\cdot \hat{u}_d||\hat{e}_{{\bf k},\sigma}\cdot \hat{u}^{\cal D}_d|}{\omega({\bf k})}$ is a slowly varying function in the reciprocal space (particularly in the region nearby the symmetric point ${\bf k}_0$), in therefore we can express the function (\ref{chapdosd2}) as
\begin{eqnarray}
{\cal S}({\bf r},{\bf r}_a,\tau)=\gamma(\frac{a}{2 \pi})^3 \sum_{\sigma} \frac{ |\hat{e}_{{\bf k}_0,\sigma}\cdot \hat{u}_d||\hat{e}_{{\bf k}_0,\sigma}\cdot \hat{u}^{\cal D}_d|}{\omega({\bf k}_0)} \int_{1BZ}d{\bf k}  e^{-i\omega({\bf k})\tau}e^{i({\bf r}-{\bf r}_a )\cdot{\bf k}}.
\label{chapdosd22}
\end{eqnarray}
Its Laplace transform is then given by
\begin{eqnarray}
&&{\cal S}({\bf r},{\bf r}_a,\omega)=
=\gamma(\frac{a}{2 \pi})^3\sum_{\sigma}\frac{ |\hat{e}_{{\bf k}_0,\sigma}\cdot \hat{u}_d||\hat{e}_{{\bf k}_0,\sigma}\cdot \hat{u}^{\cal D}_d|}{\omega({\bf k}_0)} \int_{1BZ}d{\bf k} \frac{e^{i({\bf r}-{\bf r}_a )\cdot{\bf k}}}{i(\omega({\bf k})-\omega)}.
\end{eqnarray}

In order to perform the last integrals analytically, it is convenient to consider the parabolic dispersion relation and the limit in which the detector and the emitting atom are far away from each other.

In that way, we can make a change of variable ${\bf q}={\bf k}-{\bf k}_0$, so that we have
$\omega({\bf k})=\omega_c + {\mathcal{A}} {\bf q}^2$ and extend the limits of integration to infinity. Then
\begin{eqnarray}
{\cal S} ({\bf r},{\bf r}_a,\omega)&=&-i Q \int\int\int d^3 q \frac{e^{i {\bf q}\cdot {\bf d}}}{\omega(q)-\omega}=-Q\frac{2\pi}{i d}\int_0^\infty dq q
\frac{e^{i k d}-e^{-i k d}}{\omega(q)-\omega}
\label{lucas}
\end{eqnarray}
We have also assumed that ${\bf q}=q(\sin\theta \cos\phi,\sin\theta \sin\phi,\cos\theta)$, and the vector ${\bf d}= {\bf r}-{\bf r}_a $ being parallel to the $z$ axis. The constant
\begin{eqnarray}
Q=\gamma(\frac{a}{2 \pi})^3   e^{i {\bf k}_0 \cdot {\bf d}} (1-\frac{({\bf k}_0 \cdot {\bf u}_d)^2}{k_0^2} )
(1-\frac{({\bf k}_0 \cdot {\bf u}^{\cal D}_d)^2}{k_0^2} ),
\end{eqnarray}
provided that the $d_{12}^{\cal D}=d_{12}$. Considering that $\theta$ and $\theta^{\cal D}$ are the angles between ${\bf k}_0$ and ${\bf u}_d$ and ${\bf u}^{\cal D}_d$ respectively, we have simply
\begin{eqnarray}
Q=\gamma(\frac{a}{2 \pi})^3   e^{i {\bf k}_0 {\bf d}} \sin^2 \theta \sin^2 \theta^{\cal D}
\label{cuca}
\end{eqnarray}

Notice that in our case, since we are dealing with a cubic lattice, only one symmetric point ${\bf k}_0$ and the dispersion relation around it is being considered. In a more general case, the band edge is associated with a finite collection of symmetry-related point ${\bf k}^i_0$, so that the calculus has to be taken into account the dispersion relation near each of the points, $\omega({\bf k})=\omega_c +{\mathcal{A}} ({\bf k}-{\bf k}^i_0)^2$. In such a case, the constant
\begin{eqnarray}
Q=\gamma(\frac{a}{2 \pi})^3  \sum_i  e^{i {\bf k}^i_0 {\bf d}} \sin^2 \theta_i \sin^2 \theta_i^{\cal D}
\label{cuca2}
\end{eqnarray}
Performing the last integral of (\ref{lucas}), we just obtain
\begin{eqnarray}
{\cal S} ({\bf r},{\bf r}_a,\omega)=\frac{Q 2 \pi^2}{i d {\mathcal A}}e^{-d/l}
\label{caniso2}
\end{eqnarray}
where $l=1/\sqrt{\frac{\omega-\omega_c}{{\mathcal{A}}}}$.
\footnote{
In fact the result can be expressed in the following way: In the case that $\omega_c < \omega$ then tha integral is proportional to $\exp(-i d/l)$. On the other case $\omega_c > \omega$ the integral is proportional to $\exp(-d/l)$. In both cases $l=\sqrt{{\mathcal{A}}/(|\omega_c - \omega|)}$.}

\begin{figure}[htbp]
\centerline{\includegraphics[width=0.45\textwidth]{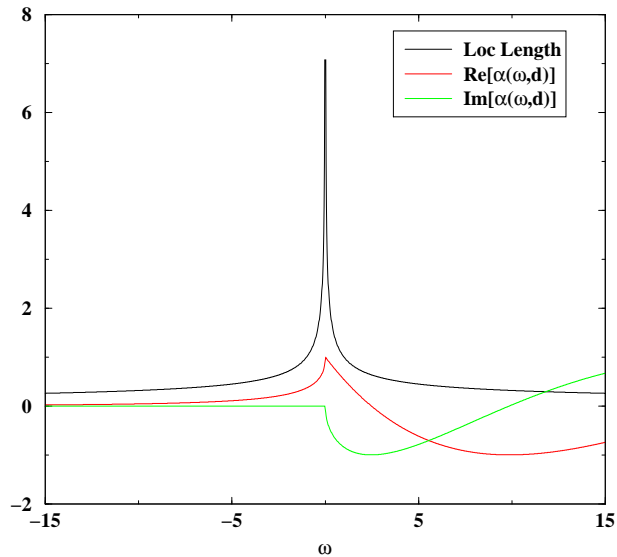}}
\caption{Real and Imaginary parts of the function (\ref{caniso2}), where we have chosen $\omega_c =0$, and ${\cal A}=1.$. The localization length is also displayed for the same parameters, showing a decreasing for increasing detunings from the band edge.  \label{locfield}}
\end{figure}

As displayed in Figure (\ref{locfield}), an increasing distance in frequencies with respect to the band edge (placed at $\omega_c =0$), gives rise to a decreasing localization length $l$, and therefore to a decreasing value of the function ${\cal S}$ \footnote{We call $l$ localization length, but note that this name is only precise when we are dealing with frequencies within the gap, and localization occurs physically.}.

\subsection{Numerical results}

In this section we compute the florescence spectra of a two-level atom placed as an impurity in a photonic crystal. To this order, we will study two cases, the florescence spectra when no spatial dependency is considered, equation (\ref{pnoesp}), and when the spatial dependency is considered (i.e. the distance from the emitting atom to the detector), in equation (\ref{pesp}). In both cases, we need to compute the Laplace transform of the system fluctuations $\langle L^\dagger(\tau)L (\tau')\rangle$, obtained with equation (\ref{qrt5}) by using the coupling operator $L$ given by (\ref{Llaser}). We show that the spatial dependency should be considered when part of the radiation is emitted in the form of evanescent waves, since in that case the approximation $\exp{(i{\bf k}\cdot{\bf r})}\approx 1$ is no longer appropriate. Evanescent modes are emitted for frequencies within the gap.

\subsubsection{No spatial dependency considered}
\label{nose}

In this section we compute the atomic fluorescence spectra according to equation (\ref{pnoesp}), by using the correlation function (\ref{caniso1}). As we will see in the next section, the formula (\ref{pnoesp}) is suitable to compute the emission spectra corresponding to frequencies outside of the gap.

\begin{figure}[htbp]
\centerline{\includegraphics[width=0.45\textwidth]{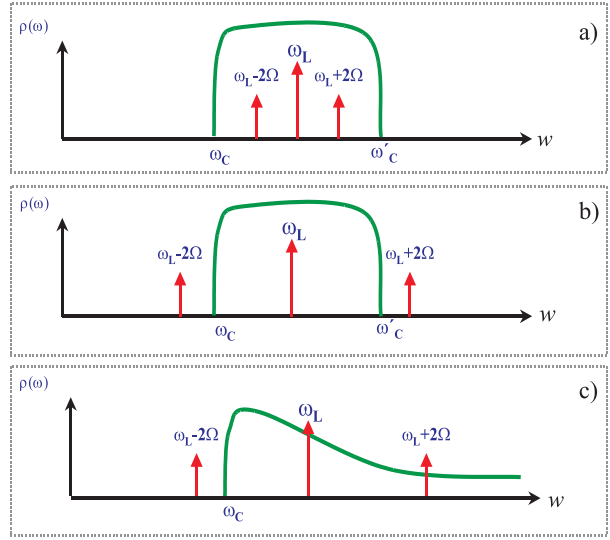}}
\caption{We display the different arrangements we consider. Figures a) and b) corresponds to the two setup described in Section (\ref{nose}), while Fig. c) corresponds to the setup of Section (\ref{sise}). \label{paper2}}
\end{figure}

\begin{figure}[htbp]
\centerline{\includegraphics[width=0.5\textwidth]{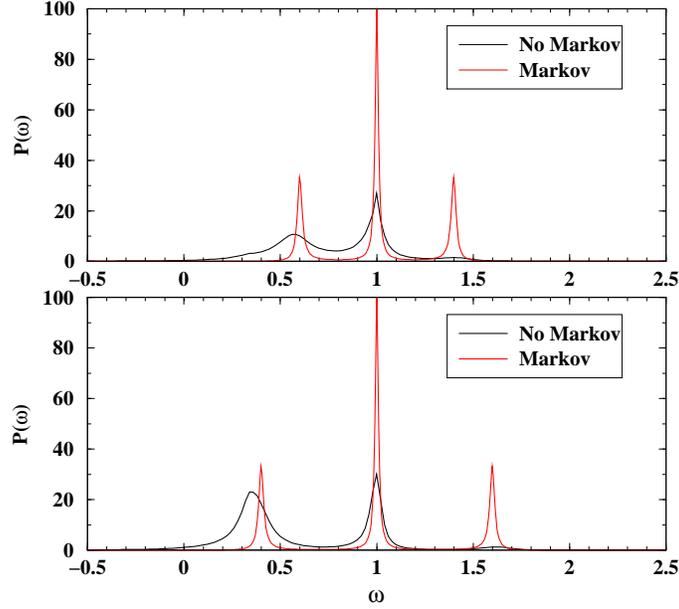}}
\caption{A comparison between the Mollow triplet obtained with Markovian interaction and the triplet obtained with the non-Markovian interaction within a PC is here shown. The tree components of the triplet are placed within the band, that corresponds to frequencies $A-B<\omega <A+B$, where $A=1$ and $B=1$. The assymetry between the left and the right side-bands described in \cite{KL04} is also described with our model.\label{laserband}}
\end{figure}

Figure (\ref{laserband}) shows a case in which we have chosen the Rabi and the laser frequencies ($\Omega$ and $\omega_L$) in such a way that the  we have placed the three components of the Mollow triplet inside the band. This corresponds to the setup of Figure (\ref{paper2}-a)). The appearance of a three component fluorescence spectra for a driven atom is due to the dressing of the atomic levels by the laser \cite{QuantumOptics}.

\begin{figure}[htbp]
\centerline{\includegraphics[width=0.5\textwidth]{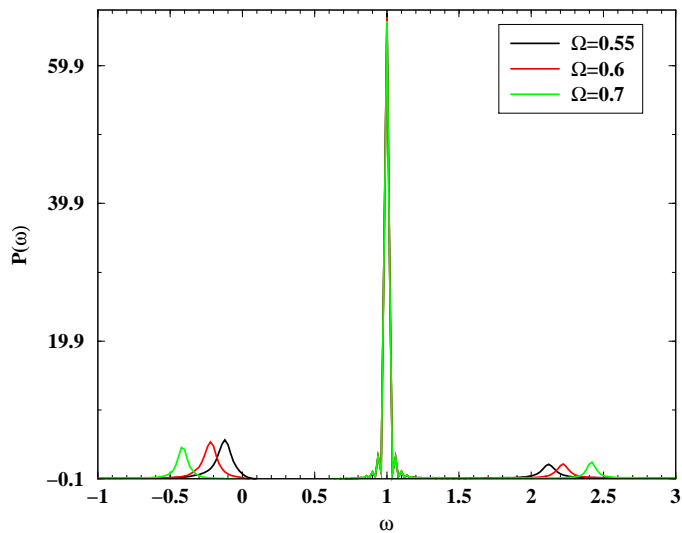}}
\caption{In contrast with Fig. (\ref{laserband}), now we place the left and right emission lines within the gap. When the distance between atom and detector is not considered, there is no filtering of the fluorescent emission that occurs within the gap.\label{lasergap}}
\end{figure}

In Figure (\ref{lasergap}) we show a case in which the two side-bands of the triplet are placed inside the gap (See Fig. (\ref{paper2}-b)). As we will clearly show in the next section, this two lines corresponds to emission of evanescent modes, which have purely imaginary wave vectors. In consequence, the approximation  $\exp{(i{\bf k}\cdot{\bf r})}\approx 1$ over estimates their magnitude, and a more accurate description of the spatial dependency is needed. Only with such description we will be able to account for the fact that evanescent modes should not be detected in the far field. This is done in the next section.

\subsubsection{Spatial dependency considered}
\label{sise}
From the last section we have concluded that for a correct description of the emission from the gap, it is very important to take into account the spatial dependency, i.e. the distance of the detector from the emitting atom. To this order, we consider the equation (\ref{pesp}), and the Laplace transform of the spatial correlation function (\ref{caniso2}) that is obtained when considering the parabolic dispersion relation $\omega({\bf k})=\omega_c +{\cal A}({\bf k}-{\bf k}_0)^2$. Moreover, in order to compute the system fluctuations (\ref{qrt5}), we take the (not spatially dependent) correlation function corresponding to this dispersion relation, which according to  \cite{FS64} is
\begin{eqnarray}
\alpha(\tau) \cong \beta^{1/2}\frac{e^{i[\pi/4-\omega_{c} \tau]}}{\tau ^{3/2}}.
\label{aniso1}
\end{eqnarray}
This correlation is just (\ref{caniso1}) taking the long time limit, so that the constants can be related as \cite{DVAG05}
\begin{equation}
\beta^{1/2}=\frac{g_{3D}^2}{8}(\frac{6}{B})^{3/2}.
\end{equation}

\begin{figure}[htbp]
\centerline{\includegraphics[width=0.5\textwidth]{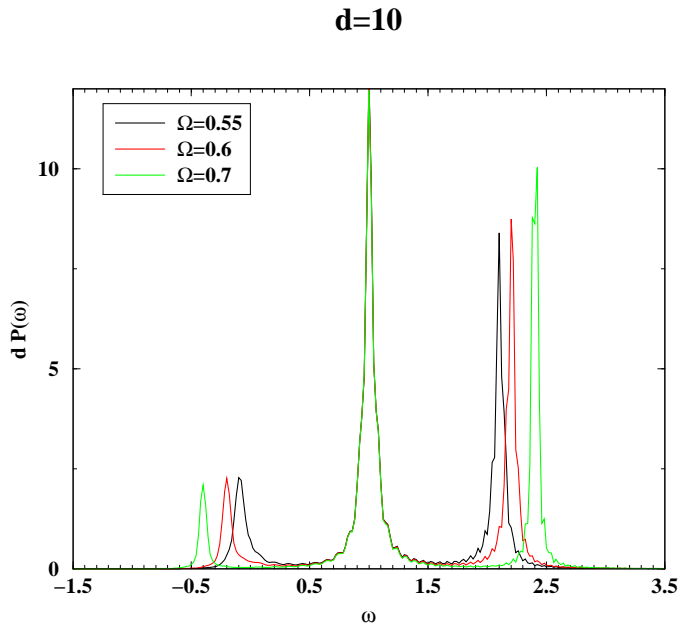}}
\caption{Power spectra $d^2 P(\omega)$ for different values of the Rabi frequency $\Omega$, and a fixed distance of the detector $d=10$. In all the cases the left sideband of the Mollow triplet is emitted in gap frequencies. While the distance $d$ is small enough to detect the evanescent modes, the fact that the localization length is smaller for increasing distance to the band edge $\omega_C=0$, makes the left Mollow component be increasingly small for increasing values of $\Omega$.  \label{locfield1}}
\end{figure}

\begin{figure}[htbp]
\centerline{\includegraphics[width=0.5\textwidth]{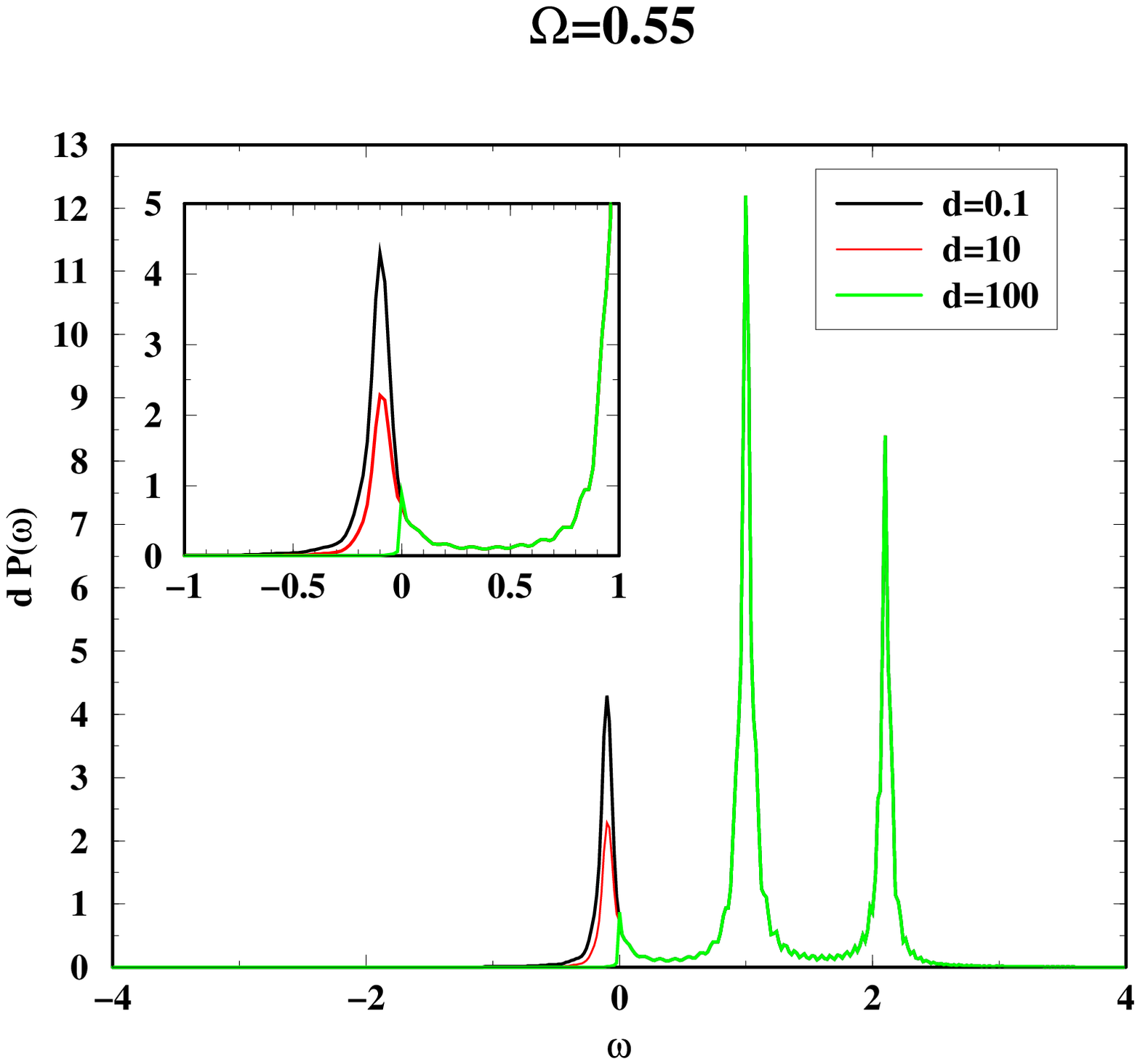}}
\caption{Power spectra $d^2 P(\omega)$ for different values of the the detector distance $d$, and fixed Rabi frequency $\Omega=0.55$, so that the left sideband of the Mollow triplet is emitted in gap frequencies. It is observed that for distances large enough this line is no longer detected, an a complete filtering of the spectra within the gap is obtained. \label{locfield2}}
\end{figure}

When using the parabolic dispersion relation, only two frequency regions have to be taken into account: a gap, for $\omega<\omega_c$, and a band for $\omega>\omega_c$. We chose our parameters ($\Delta_{AL}=0$, $\omega_L =1$, and $\Omega$) such that the left Mollow component is placed in the gap, while central and right component have frequencies within the band (See Fig. (\ref{paper2}-c) for a schematic view.). In figure (\ref{locfield1}) we show the fluorescence spectra for a certain distance $d$ of the detector (expressed by taking the periodicity of the crystal as the unity) which is small enough as to be still detecting some evanescent modes from the gap. Different choices of $\Omega$ shows that the more deep in the gap the emission occurs, the less we are efficient to detect it, due to the decreasing of the localization length. Figure (\ref{locfield2}) shows the spectra for a fixed value of $\Omega$, but varying the distance of the detector, in such a way that for $d$ large enough, a complete suppression of fluorescence from the gap (corresponding to the left side-band) occurs.

Concerning this, two important remarks have to be made here.
\begin{itemize}
\item In contrast to \cite{VJB02} and \cite{KL04}, where a filter $\Theta(\omega-\omega_c)$ is introduced for emission frequencies within the gap, in the description presented here no filtering effect occurs unless the spatial dependency is considered, and the detector is placed in the far field.
\item A second remark concerns the difference between the spontaneous and the fluorescence spectra. Even considering the spatial dependency and the detector in the far field, some radiation is detected within the gap region in the spontaneous emission spectra. The reason is that there is a transient field, called diffusive field, that appears at the beginning of the atomic evolution. Since this field can be emitted in the gap frequencies, it carries away energy from the atom. Hence, it is responsible of the slight dissipation that occurs in the atomic population of the excited level at the initial time steps of the evolution \cite{FS64,DVAG05}. A description of this field, including its spatial dependency is found in \cite{YY00}. In this paper, this is done by solving the atom dynamics within the one photon sector \footnote{Note that this procedure cannot be used here, since in the dynamics of a driven atom more than one photon come into play.}.

In contrast, the fluorescence spectra deals with the signal emitted by the atom in the stationary limit, and therefore the transient field is no longer detected. For that reason, the radiation within the gap is only composed of evanescent modes, and cannot be observed when placing the detector far away from the atom.

In summary, it is true that the filtering appears as a consequence of the special form of the dispersion relation of the field in the PC. However, we show here that this filtering only occurs provided that the signal is in the stationary limit, and the distance between the emitting source and the detector is large.

\end{itemize}

\section{Conclusions}
We present in this paper a general formula to compute the emission spectra of an atom linearly coupled to an environment with non-Markovian interaction, and within the weak coupling approximation.
This formula is explicitly dependent on the distance between the emitting atom and the detector, ${\bf d}$.
We apply it to compute the fluorescence spectra of a two-level atom in a photonic crystal. In those materials, the electromagnetic field have very specific features, showing a gap or range of frequencies where the propagating modes are not allowed. In that frequency region, the emission spectra depends critically on ${\bf d}$, since the emission occurs in the form of evanescent (non propagating) modes, which decay exponentially from the atom. For that reason, we show that in the far field these modes are not detected.
We note that the description of this evanescent field can only be made with a non-Markovian description, since a Markov theory, with dissipation rates calculated through the Fermy Golden Rule, predicts no emission at all within the gap region.


\section*{Acknowledgements} We thank H. Carmichael, G.C. Hegerfeldt, A. Ru{\'i}z and L. S. Schulman for their comments and G. Nicolis. P. Gaspard, J.I. Cirac and W.T. Strunz for support and encouragement. We thank R.F. O'Connell for pointing out some important references concerning the Quantum Regression Theorem. This work has been supported by the Gobierno de Canarias (Spain) (PI2004/025) and Ministerio de Ciencia y Tecnolog\'\i a of Spain (FIS2004-05687,FIS2007-64018). I. de Vega is financially supported by a Ministerio de Ciencia y Tecnolog\'\i a (EX2006-0295).

\end{document}